\documentclass[abstract=on,notitlepage,superscriptaddress,nofootinbib,aps,showpac,prd]{revtex4}
\usepackage{amsfonts}
\usepackage{amsmath}
\usepackage{amssymb}
\usepackage{enumerate}
\usepackage{graphicx}
\usepackage{color}
\usepackage{epstopdf}
\usepackage{float}
\usepackage{dcolumn}
\usepackage{hyperref}
\usepackage{amsthm}

\begin{document}


\title{The interaction between gravitational waves and a viscous fluid shell on a Schwarzschild background}



\author{Nigel T. Bishop}\email[]{n.bishop@ru.ac.za}
\affiliation{Department of Mathematics, Rhodes University}

\begin{abstract}
Previous work has shown that the interaction between gravitational waves (GWs) and a shell of viscous matter leads to damping of the GWs and heating of the matter, and that these effects may be astrophysically significant. This result was derived using the theory of linear perturbations about a Minkowki background, and in this work the model is extended to be more physically realistic by allowing the background geometry to be Schwarzschild. It is found that the difference between using a Schwarzschild or Minkowski background is minimal when either $r\gg M$ or $\lambda< M$, where $r$ is the radius of the shell, $\lambda$ is the GW wavelength and $M$ is the mass of the system in geometric units (so that $1M_\odot=1.48$km). However, when $r\sim 6M$ and $\lambda\sim 25M$, then the damping and heating effects are about 9 times larger on a Schwarzschild background than on Minkowski, and such situations occur astrophysically.

	\end{abstract}
	
	\maketitle
	
	
	\section{Introduction}
	\label{intro}

Gravitational waves (GWs) travelling through a perfect fluid do not experience any absorption or dissipation~\cite{Ehlers1987} as noted in~\cite{Lu:2018smr}. However, Hawking ~\cite{Hawking1966} showed that in the case of nonzero shear viscosity, $\eta$, GWs travelling through such a fluid would lose energy to the medium so that the GWs would be damped; see also~\cite{Madore1973,Prasanna1999,Goswami2017}. In a series of recent papers, we have shown how a dust shell surrounding a GW event modifies the GWs in both magnitude and phase~\cite{Bishop:2019eff,Naidoo:2021}; and that if the matter in the shell is viscous then GWs are damped~\cite{bishop2022effect} and the matter heated~\cite{bishop2024heating,bishop2024heating2}. A key point about these effects is that they can be large and astrophysically significant as $\lambda/ r$ increases, where $\lambda$ is the GW wavelength and $r$ is the radius of the shell. 

A limitation of our previous work is that it uses perturbations about a Minkowski background; however, a Schwarzschild background would be more appropriate for the astrophysical scenarios that were considered. This paper develops the theory of the interaction between GWs and a viscous shell of matter on a Schwarzschild background. The extension from the background being Minkowski to Schwarzschild is not straightforward: in the Minkowski case linearized perturbations are described by simple analytic formulas, whereas in the Schwarzschild case part of the solution has to be found numerically.

	The paper uses the Bondi-Sachs formalism, and Sec.~\ref{theory} presents previous results about linearized perturbations on both Minkowski and Schwarzschild backgrounds. Then Sec.~\ref{s-metric} describes numerical procedures that calculate the metric perturbations. Sec.~\ref{s-matter} presents formulas for the velocity field from which the shear tensor $\sigma_{ab}$ is calculated leading to the rate of energy transfer from the GWs to the matter. Sec.~\ref{s-results} gives the numerical results in two parts: First, Sec.~\ref{s-code} gives the errors found for various code validation tests, and then Sec.~\ref{s-shear} presents numerical results comparing the GW damping and heating effects when the background is changed from Minkowski to Schwarzschild. These results are discussed in Sec.~\ref{s-conclusion}. The linearized Einstein equations are given in Appendix~\ref{a-Ee}; and the computer scripts used are described in Appendix~\ref{a-MapleOctave}, with the scripts available as Supplemental Material.

	We use geometric units in this paper, with the gravitational constant $G$ and the speed of light $c$ set to unity. 
	
	\section{Linearized perturbations in the Bondi-Sachs formalism}
	\label{theory}
	
	We use the formalism developed previously~\cite{Bishop:2019eff,Naidoo:2021} on GWs propagating through matter shells. The metric is in Bondi-Sachs form~\cite{Bondi62,Sachs62}
	\begin{align}
		ds^2  = & -\left(e^{2\beta}\left(1 + rW_c\right)
		- r^2h_{AB}U^AU^B\right)du^2
		- 2e^{2\beta}dudr \nonumber \\
		& - 2r^2 h_{AB}U^Bdudx^A
		+  r^2h_{AB}dx^Adx^B\,,
		\label{eq:bmet}
	\end{align}
	where $h^{AB}h_{BC}=\delta^A_C$, and the condition that $r$ is a surface area coordinate implies $\det(h_{AB})=\det(q_{AB})$ where $q_{AB}$ is a unit sphere metric (e.g. $d\theta^2+\sin^2\theta d\phi^2$). We represent $q_{AB}$ by a complex dyad (e.g. $q^A=(1,i/\sin\theta)$) and introduce the complex differential angular operators $\eth,\bar{\eth}$~\cite{Newman-Penrose-1966}, with the operators defined with respect to the unit sphere as detailed in~\cite{Bishop2016a,Gomez97}. Then $h_{AB}$ is represented by the complex quantity $J=q^Aq^Bh_{AB}/2$ (with $J=0$ characterizing spherical symmetry), and we also introduce the complex quantity $U=U^Aq_A$.
	
The Eddington-Finkelstein form of the Schwarzschild metric is Eq.~\eqref{eq:bmet} with
\begin{equation}
W_c=-\frac{2M}{r^2}\,,\beta=J=U=0\,,
\end{equation}
and we make the ansatz of a small perturbation about Schwarzschild spacetime with the metric quantities $\beta,U,W_c,J$ taking the form
	\begin{align}
		\beta=&\Re(\beta^{[2,2]}(r)e^{i\nu u}){}_0Z_{2,2}\,,\;\;
		U=\Re(U^{[2,2]}(r)e^{i\nu u}){}_1Z_{2,2}\,,\nonumber \\
		W_c&=-\frac{2M}{r^2}+\Re(W_c^{[2,2]}(r)e^{i\nu u}){}_0Z_{2,2}\,,\;\;
		J=\Re(J^{[2,2]}(r)e^{i\nu u}){}_2Z_{2,2}\,.
		\label{e-ansatz}
	\end{align}
	The perturbations oscillate in time with frequency $\nu/(2\pi)$. The quantities ${}_s Z_{\ell,m}$ are spin-weighted spherical harmonic basis functions related to the usual ${}_s Y_{\ell,m}$ as specified in~\cite{Bishop-2005b,Bishop2016a}. They have the property that ${}_0 Z_{\ell,m}$ are real, enabling the description of the metric quantities $\beta,W$ (which are real) without mode-mixing; however, for $s\ne 0$ ${}_s Z_{2,2}$ is, in general, complex. A general solution may be constructed by summing over the $(\ell,m)$ modes, but that is not needed here, since we are considering a source that is continuously emitting GWs at constant frequency dominated by the $\ell=2$ (quadrupolar) components. The resulting linearized vacuum Einstein equations are given in Appendix~\ref{a-Ee}, Eqs.~\eqref{E-11} to \eqref{E-0A}.

It was shown previously~\cite{Bishop-2005b} that manipulation of Eqs.~\eqref{E-1A} and \eqref{E-J} leads to a master equation
\begin{equation}
-2 J_2( 2x +8Mx^2 +i\nu) +2\frac{dJ_2}{dx}\left(2x^2+i\nu x
-7x^3 M\right) +x^3(1-2xM)\frac{d^2J_2}{dx^2}=0
\label{e-S}
\end{equation}
where $J_{2}(x)\equiv d^2J^{[2,2]}/dx^2$ and $x=1/r$. In the Minkowski background case $M=0$, Eq.~\eqref{e-S} has a simple analytic solution
\begin{equation}
J_2(x)=b_1x+b_2 \exp\left(\frac{2i\nu}{x}\right)
\left(3x-6i\nu -\frac{6\nu^2}{x}+\frac{4i\nu^3}{x^2}+\frac{2\nu^4}{x^3} \right)\,,
\end{equation}
where $b_1,b_2$ are integration constants; further, if there are no incoming GWs, then $b_2=0$ and the various metric terms $J^{[2,2]}(r)$ simplify to polynomials in $1/r$. {\bf However}, in the Schwarzschild case $M\ne 0$, it is not possible to obtain an algebraic expression for the solution of Eq.~\eqref{e-S}, and construction of a numerical solution is not straightforward since the differential equation has an essential singularity at $x=0$.
Previous work~\cite{Bishop2009} (see also~\cite{Mongwane2024}) has constructed a numerical solution of Eq.~\eqref{e-S}, within the context of a study of Schwarzschild quasinormal modes, and we now outline that procedure.

A series solution $\sum_{n=1}^N a_nx^n$ about $x=0$ is generated by the recurrence relation
\begin{equation}
a_n=-a_{n-1}\frac{n^2+n-6}{2i\nu (n-1)}+a_{n-2}M\frac{2n(n+2)}{2i\nu(n-1)}
\end{equation}
with $a_1=1,a_2=0$. This solution is normalized to have  $\lim_{x\rightarrow 0} J_2(x)/x=1$ and will be denoted by $J_{2T}$. Although the series is not convergent, it is possible to determine a bound on the error~\cite{Olver74}. More precisely, given $\epsilon>0$, values for  $x_0,N$ can be determined such that
\begin{equation}
\left\vert J_{2T}(x)-\sum_{n=1}^N a_n x^n\right\vert <\epsilon\;\;
\forall x\;\mbox{ with }0<x\le x_0\,,
\label{e-an}
\end{equation}
Thus, values for $J_{2T}(x_0),\partial_x J_{2T}(x_0)$ are determined which are used as initial data for a numerical solution of Eq.~\eqref{e-S}. Choosing $\epsilon=10^{-16}$, i.e. machine precision, means that the error in the initial data does not affect the accuracy of the numerical solution to be obtained.

Eq.~\eqref{e-S} is recast into Ricatti form by means of the transformation
\begin{equation}
V(x)=x^2 \frac{\partial_x J_{2T}(x)}{J_{2T}(x)}
\end{equation}
leading to
\begin{align}
\frac{dV}{dx}&=\frac{1}{x^2(-2Mx)}\times\nonumber \\
&\left(4x^2-2Vx-V^2+16Mx^3+10x^2MV+2i\nu x-2i\nu V+2MV^2x\right)\,.
\label{e-deV}
\end{align}
The parameter $M$ is now set to $M=1$ (Schwarzschild case) or $M=0$ (Minkowski case), and Eq.~\eqref{e-deV} is solved numerically over the range $x_0\le x\le 0.25$. Eq.~\eqref{e-S} has a regular singularity at the black hole horizon $x=0.5$, and since we do not need the solution near the horizon, the solution is constructed only for $x\le0.25$, i.e. $r\ge 4$. Then
\begin{equation}
J_2(x)=12\sqrt{6}\left(J_{2T}(x_0)+\exp\left(\int_{x_0}^x \frac{V(s)}{s^2}ds\right)\right) \,,
\end{equation}
and then $\partial_x J(x)$ and $J(x)$ are obtained by numerical integration; the factor $12\sqrt{6}$ is for later convenience. Substituting $x=1/r$ gives $J^{[2,2]}(r)$, and straightforward calculation gives
\begin{equation}
\partial_r J^{[2,2]}=-x^2\partial_x J\,,\;
\partial^2_r J^{[2,2]}=x^3(2\partial_x J+x\partial^2_x J)\,.
\end{equation}
The manipulation of Eqs.~(\ref{E-1A}) and (\ref{E-J}) to give Eq.~\eqref{e-S} also leads to
\begin{align}
U(x)=&0.5(2x^2\partial_x J-i\nu x^2\partial^2_xJ
+x^5M\partial^3_xJ-x^3\partial_x^2J+2x^3M\partial_xJ
+4x^4M\partial_x^2J/2\nonumber \\
&-x^4\partial_x^3J+i\nu x\partial_xJ-i\nu J)\,,
\end{align}
and substituting $x=1/r$ gives $U^{[2,2]}(r)$.
Eq.~\eqref{E-AB} provides an expression for $\partial_r(W_c^{[2,2]}r^2)$, and integration then leads to $W_c^{[2,2]}$.

\section{Computation of the metric coefficients}
\label{s-metric}
Solving the Einstein equations  leads to a solution involving a number of constants of integration, some of which represent gauge freedoms and so are freely specifiable. For numerical work, these constants need to be specified, and we use the ``Bondi'' gauge in which the metric is explicitly asymptotically flat, i.e. all metric terms fall off to the Minkowski values at least as fast as $1/r$. This implies that the constant of integration in the solution to Eq.~\eqref{E-11} is set to $0$ so that
\begin{equation}
\beta^{[2,2]}(r)=0\,.
\end{equation}
Further, $J^{[2,2]}(r)$ must satisfy $\lim_{r\rightarrow\infty} J^{[2,2]}(r)=0$.

	As shown in previous work~\cite{Bishop-2005b,Bishop:2019eff}, solving the vacuum Einstein equations under the condition of no incoming radiation leads to
	\begin{align}
		\beta^{[2,2]}=&0\,, \nonumber \\
		W_c^{[2,2]}=&-\frac{12  C_{50}}{r^2}
		+C_{40}f_W(r) 	\,,\nonumber\\
		U^{[2,2]}=&\frac{2\sqrt{6} C_{30}}{r^2}+C_{40}f_U(r)
		\,,\nonumber \\
		J^{[2,2]}=&\frac{2\sqrt{6}C_{30}}{r}
		+C_{40}f_J(r)\,,
		\label{e-pert}
	\end{align}
where the functions $f_W(r),f_U(r),f_J(r)$ depend on the paramaters $\nu,M$ and in the Minkowski $M=0$ case are
\begin{equation}
f_W=-6\left(\frac{2i\nu}{r^3}+\frac{1}{r^4}\right)\,,\;f_U=-\sqrt{6}\left(\frac{3}{r^4}+\frac{4i\nu}{r^3}\right)\,,\;f_J=\frac{2\sqrt{6}}{r^3}\,.
\label{e-pertM}
\end{equation}
The constraints, Eqs.~\eqref{E-00} to \eqref{E-0A}, provide two conditions on the constants $C_{30}$, $C_{40}$ and $C_{50}$. In the Minkowsi case, substitution of Eqs.~\eqref{e-pert}, \eqref{e-pertM} into the constraints leads to simple algebraic condions which are easily solved to give
\begin{equation}
C_{30}=-\nu^2C_{40}\,,\;C_{50}=12\nu^2C_{40}\,.
\end{equation}
However, when $M\ne 0$, the solution Eq.~\eqref{e-pert} is partly numerical and the procedure has to be amended.

The first step is to construct a numerical solution in which the integration constants $C_{30}=C_{50}=0$, and then the constraint equations are evaluated on the finite grid $x_0\le x\le 0.25$, with the values obtained denoted as $R_{00}^{[0]},R_{0A}^{[0]}$. Next, Eqs.~\eqref{e-pert} with $C_{40}=0$ are substitued into the constraints yielding algebraic expressions. The constraints are satisfied provided
\begin{align}
R_{00}:\;\;&C_{30} 6i\nu\left(\frac{2}{r^2}+\frac{3M}{r^3}\right) 
+C_{50}\left(\frac{i\nu}{r^2}-\frac{3}{r^3}\right) +R_{00}^{[0]}=0\nonumber\\
q^AR_{0A}:\;\;&C_{30} \frac{3i\nu M\sqrt{6}}{r^2} 
-C_{50}\frac{\sqrt{6}}{2r^2} +q^AR_{0A}^{[0]}=0
\label{e-C3050}
\end{align}
We now find values of $C_{30},C_{50}$ such that the sum over grid points of the squares of the error is minimized, for which there are standard procedures in numerical linear algebra. Then these values are used in Eq.~\eqref{e-pert} to obtain the solution for the metric functions.

Since the calculations here are in the Bondi gauge, the expression for the gravitational news is simply ${\mathcal N}_{0}=\sqrt{6}\nu \Re(iC_{30}\exp(i\nu u))\,{}_2Z_{2,2}$. The rescaled gravitational wave strain and news are related (see~Eq.~(276) in \cite{Bishop2016a}) $\mathcal{H}_{0}=r(h_+ +ih_\times) = 2\int{\mathcal N}_{0} du$, giving
	\begin{equation}
		{\mathcal H}_{0}=\Re(H_{0} \exp(i\nu u))\,{}_2Z_{2,2}\;\mbox{with}\;\; H_{0}=2\sqrt{6} C_{30}\,.
		\label{e-HM0}
	\end{equation}

	\section{Matter shell}
	\label{s-matter}
	
	We now suppose that the GWs pass through a shell of matter, and determine the velocity field $V_a$. The ansatz for $V_a$ is 
	\begin{align}
		V_0&=-\sqrt{1-\frac{2M}{r}}+\Re(V^{[2,2]}_0(r)e^{i\nu u}){}_0Z_{2,2}\,,\nonumber \\
		V_1&=-\frac{1}{\sqrt{1-\frac{2M}{r}}}+\Re(V^{[2,2]}_1(r)e^{i\nu u}){}_0Z_{2,2}\,,\;
		q^A V_A=\Re(V^{[2,2]}_{ang}(r)e^{i\nu u}){}_1Z_{2,2}\,.
		\label{e-Vansatz}
	\end{align}
We suppose that the background matter density is constant $\rho_0$ and a background pressure field $p(r)$ is needed to prevent matter falling towards the black hole; however, these do not appear in the formulas for the velocity. The velocity normalization condion $V_aV_bg^{ab}=-1$ gives
\begin{equation}
V^{[2,2]}_0=-W_c^{[2,2]}\frac{r}{2\sqrt{1-\frac{2M}{r}}}
\label{e-V0}
\end{equation}
	Then solving the matter conservation conditions $\nabla_a\left[(\rho+p) v^aV^b-p g^{ab}\right]=0$ leads to
\begin{equation}
V^{[2,2]}_{ang}(r)=i\frac{rW_c^{[2,2]}}{2\nu\sqrt{1-\frac{2M}{r}}}
\label{e-Vang}
\end{equation}
and
\begin{equation}
\frac{dV_1^{[2,2]}}{dr}=\frac{2M^2-2rM+iM\nu r^2-r^4\nu^2}{r^2M\left(1-\frac{2M}{r}\right)}V_1^{[2,2]}+c_{0,v}\,,
\label{e-dV1}
\end{equation}
where $c_{0,v}$ is given in the computer algebra output, see Appendix~\ref{a-MapleOctave}. Eq.~\eqref{e-dV1} is solved numerically.
	
The shear tensor of the fluid flow $\sigma_{ab}$ is defined as (e.g., see~\cite{Baumgarte2010a}, p.139) 
\begin{equation}
			\sigma_{ab}=\frac{(P_{ac}\nabla_d V_b +P_{bc}\nabla_dV_a)g^{cd}}{2}-\frac{P_{ab}\theta}{3}\,,
\end{equation}		
where the fluid expansion $\theta=g^{ab}\nabla_a V_b$ and where $P_{ab}=g_{ab}+V_a V_b$ is the projection tensor. Expressions for these quantities, in terms of metric coefficients and $V_1^{[2,2]}$ are obtained using computer algebra, see Appendix~\ref{a-MapleOctave}.
Then the rate of energy loss per unit volume is $\dot{E}=-2\eta \sigma_{ab}\sigma^{ab}$ where $\eta$ is the coefficient of shear viscosity~\cite{Baumgarte2010a}. An expression for $\dot{E}$ is obtained using computer algebra, and then integrated over a shell of radius $r$ and thickness $\delta r$; the integration is straightforward because of the orthonormality of the angular basis functions ${}_sZ_{\ell,m}$. 
We also time-average $\dot{E}$ over a wave period, which is again straightforward. Every term has time-dependence $\cos^2(\nu u)$, $\sin^2(\nu u)$ or $\cos(\nu u)\sin(\nu u)$, and the time-average of the first two cases is $1/2$ and of the last case is $0$.

Following the procedure of~\cite{bishop2022effect}, we find that
	\begin{equation}
		\left<\dot{E}_S\right> = -16\pi\eta\frac{G}{c^3} \delta r \left<\dot{E}_{GW}\right>
		f_E(r,\nu,M)\,.
	\end{equation}
where $\left<\dot{E}_S\right>$ is the time-averaged rate of energy increase to a spherical shell of radius $r$ and thickness $\delta r$, and $\left<\dot{E}_{GW}\right>$ is the time-averaged gravitational wave power passing through the shell. In the case of a Minkowski background with $M=0$, the function $f_E$ is a simple analytic expression~\cite{bishop2022effect}
		\begin{equation}
f_E(r,\nu,0)=\left(1+\frac{2}{r^2\nu^2}+\frac{9}{r^4\nu^4}
		+\frac{45}{r^6\nu^6}+\frac{315}{r^8\nu^8}
		\right)\,,
		\label{e-dE_eta}
	\end{equation}
but for $M\ne 0$ the expression is lengthy and is given in the computer algebra output, see Appendix~\ref{a-MapleOctave}.

\section{Computational results}
\label{s-results}

\subsection{Code verification}
\label{s-code}
In the Minkowski background $M=0$ case, analytic formulas are available for the metric coefficients $J,U,W_c$ (Eqs.~\eqref{e-pert} and \eqref{e-pertM}) as well as for $f_E$ (Eq.~\eqref{e-dE_eta}). Thus it is straightforward to run the Matlab/Octave code with $M=0$ and determine the maximum of the absolute value of the error for each of the quantities listed. The results are given in Table~\ref{t-Mink}
 \begin{table}[h]
 \begin{tabular}{ll}
 \hline
	Quantity &    $||\mbox{Error}||_\infty$
	\\
	\hline
	\hline
   $J$ & $5.3\times 10^{-10}$   \\
   $U$ & $1.6\times 10^{-9} $\\
   $W_c$ & $1.7\times 10^{-9}$   \\
   $f_E$ & $2.5\times 10^{-8}$ \\
	\hline
	\hline
	\end{tabular}
	\caption{Errors in the $L_\infty$ norm in the Minkowski background ($M=0$) case with $\nu=0.5$.}
	\label{t-Mink}
	\end{table}

In the Schwarzschild $M=1$ case, the numerical solutions obtained are substituted into the left hand side of the Einstein equations, Eqs.~\eqref{E-1A} to \eqref{E-0A}, and we report the maximum of the absolute value in Table \ref{t-Gen}.
 \begin{table}[h]
 \begin{tabular}{ll}
 \hline
	Quantity &    $||\mbox{Error}||_\infty$
	\\
	\hline
	\hline
   $q^AR_{1A}$ & $7.8\times 10^{-15}$   \\
   $h^{AB}R_{AB}$ & $7.5\times 10^{-15} $\\
   $q^Aq^BR_{AB}$ & $4.1\times 10^{-14}$   \\
   $R_{01}$ & $4.7\times 10^{-16}$ \\
   $R_{00}$ & $6.8\times 10^{-10}$ \\
   $q^AR_{0A}$ & $6.4\times 10^{-10}$ \\
	\hline
	\hline
	\end{tabular}
	\caption{$L_\infty$ norm errors in the Einstein equations for the Schwarzschild background ($M=1$) case with $\nu=0.5$.}
	\label{t-Gen}
	\end{table}

\subsection{Effect of $M\ne 0$ on magnitude of the fluid shear}
\label{s-shear}
\begin{figure}
\includegraphics[width=15cm]{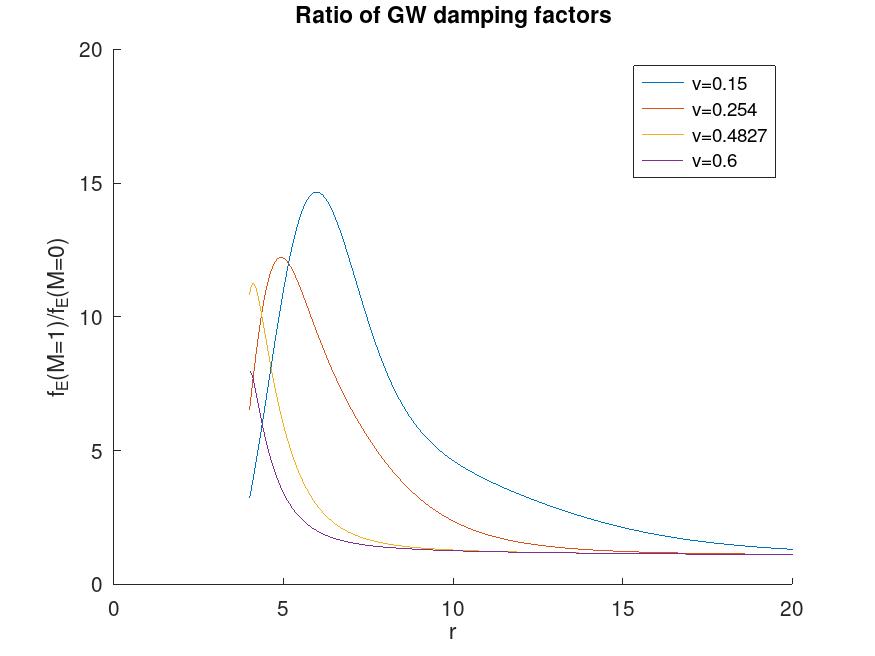}
\caption{Plots of $f_E(r,\nu,1)/f_E(r,\nu,0)$ against $r$ for values of $\nu$ shown}
\label{f-TI}
\end{figure}

In Fig.~\ref{f-TI}, we plot $f_E(r,\nu,M)/f_E(r,\nu,0)$ for a range of values of $\nu$. The choice of values is guided by astrophysical considerations. GW150914 had a remnant with mass $M=62M_\odot=62\times 1480$m, and a frequency that increased during inspiral to $132$Hz at merger and $251$Hz during ringdown. Using the formula
\begin{equation}
\nu=\frac{2\pi f}{c} M=\frac{2\pi}{\lambda}M
\end{equation}
where $f$ is the frequency, $\lambda$ is the wavelength and the speed of light $c=2.998\times 10^8$m/s, we find $132$Hz$\,\rightarrow \nu=0.254, M/\lambda=0.0404$ and $251$Hz$\,\rightarrow \nu=0.483, M/\lambda=0.0768$. Since our computations scale the mass of the gravitating system to $M=1$, these values for $\nu$ would apply to any black hole merger.

Fig.~\ref{f-TI} shows that using a background of Minkowski rather than of Schwarzschild underestimates the GW damping and heating effects, although the difference becomes negligible as $r/M$ becomes large. Further, the magnitude of this effect reduces as the wave frequency increases. However, it is also clear that there are cases where use of a Minkowski background leads to a significant underestimate of the effect. For example, use of a Minkowski background underestimates the damping/heating effect by a factor of $9$ for matter at $r=6M$ at the time of merger of two black holes with $\nu=0.254, M/\lambda=0.0404$.

\section{Conclusion}
\label{s-conclusion}

This paper has extended previous results on the interaction bewteen GWs and viscous matter from a background spacetime of Minkowski to one where the background geometry is Schwarzschild. Perhaps surprisingly, the results in Fig.~\ref{f-TI} show that the GW damping and heating effects are enhanced by a factor of order 10 for matter close to the gravitating mass and at lower frequencies.  This factor is rather large, and so at a heuristic level we investigated its origin. Firstly, at $r=6M$ the metric coefficient $J^{[2,2]}$ is about 9\% smaller than in the Minkowski case, and $U^{[2,2]},W_c^{[2,2]}$ are up to 2\% larger (see Appendix~\ref{a-MapleOctave}), so the change in their values makes only a minimal contribution to the effect. However, defining $S=1-2M/r$ and noting that $S=2/3$ at $r=6M$, Eqs.~\eqref{e-Vang} and \eqref{e-Vang} show that $V^{[2,2]}_{0}$ and $V^{[2,2]}_{ang}$ increase by $\sqrt{1/S}$. Further, $V^{[2,2]}_{1}$ (found numerically, see Appendix~\ref{a-MapleOctave}) increases by a factor of $4.02$. Also, the formula for $f_E$ (see Appendix~\ref{a-MapleOctave}) contains velocity-squared terms, as well as terms multiplied by $S^{-3}=27/8$. Thus, the increase by a factor of $9$ is not unreasonable.

The numbers used in the example discussed in Sec.~\ref{s-shear} were motivated by astrophysical considerations, thus it is clear that the theory developed here is relevant to astrophysics. However, the scope of this paper is limited to theory and computational tools, and its application to various astrophysical scenarios will be explored in future works.

	%
	
	\appendix
	\section{Einstein equations}
		\label{a-Ee}
The vacuum Einstein equations linearized about Schwarzschild when the metric coefficients are given by Eq.~\eqref{e-ansatz}, and subject to the gauge condition $\beta^{[2,2]}(r)=0$, are
\begin{align}
R_{11}:\;\; &\partial_r\beta^{[2,2]}=0 \,,
\label{E-11}\\
q^AR_{1A}:\;\; &2\partial_rJ^{[2,2]}-4r\partial_rU^{[2,2]} -r^2\partial_r^2U^{[2,2]}=0\,,
\label{E-1A}\\
h^{AB}R_{AB}:\;\; &\partial_r\left({W_c^{[2,2]}}{r^2}\right)-\sqrt{6}J^{[2,2]} +2\sqrt{6}rU^{[2,2]}
  +\frac{\sqrt{6}}{2}r^2\partial_rU^{[2,2]}=0\,,
  \label{E-AB}\\
q^Aq^BR_{AB}:\;\;&(2rM-r^2)\partial_r^2J^{[2,2]}+(2M-2r+2i\nu r^2)\partial_rJ^{[2,2]}
  +2i\nu rJ^{[2,2]} \nonumber \\
  &+2r^2\partial_r U^{[2,2]}+4rU^{[2,2]} =0\,,
  \label{E-J}\\
R_{00}:\;\;&\frac{6W_c^{[2,2]}}{r} -2i\nu W_c^{[2,2]}-2\sqrt{6}i\nu U^{[2,2]}
  -2\sqrt{6}\frac{MU^{[2,2]}}{r^2}\nonumber \\
   &-\frac 1r  \left(1-\frac{2m}{r}\right)\partial_r^2\left({W_c^{[2,2]}}{r^2}\right)=0\,,
   \label{E-00}\\
R_{01}:\;\; &\frac 1r\partial_r^2\left({W_c^{[2,2]}}{r^2}\right)+\sqrt{6}\partial_rU^{[2,2]}
+2\sqrt{6}\frac{U^{[2,2]}}{r}=0\,,
\label{E-01}\\
q^AR_{0A}:\;\; &\sqrt{6}W_c^{[2,2]}-\frac{\sqrt{6}}{r}\partial_r\left({W_c^{[2,2]}}{r^2}\right)+i\nu r^2\partial_rU^{[2,2]}-2U^{[2,2]}-4r\partial_r U^{[2,2]}\nonumber \\
  &-r^2 \partial_r^2 U^{[2,2]}+8M\partial_rU^{[2,2]}+2Mr\partial^2_rU^{[2,2]}+2i\nu J^{[2,2]}=0\,.
  \label{E-0A}
\end{align}

	\section{Computer algebra and numerical scripts}
	\label{a-MapleOctave}

The computer scripts used in this paper are written in plain text format, and are available as Supplementary Material. Output files should be viewed using a plain text editor with line-wrapping switched off.
	
	The file driving the calculation is \texttt{SchwarzschildShell.map}, which takes input from the files \texttt{gamma.out, initialize.map, lin.map} and \texttt{ProcRules.map}; the output is in \texttt{SchwarzschildShell.out}. The Maple script is adapted from those reported in our previous work~\cite{Bishop:2019eff,bishop2022effect}. The stress-energy conservation condition $\nabla_aT^{ab}=0$ is used, together with Eqs.~\eqref{e-ansatz} and \eqref{e-Vansatz}, to find the velocity components in Eqs.~\eqref{e-V0} and \eqref{e-Vang} and the coefficients in Eq.~\eqref{e-dV1}. Next, the shear tensor $\sigma_{ab}$ is evaluated, leading to the quantity $f_E$ in Eq.~\eqref{e-dE_eta}; the script also translates this output into Matlab/Octave form in the file \texttt{fE.out}. Finally, the script calculates the Einstein equations presented in Appendix~\ref{a-Ee}.

The Matlab/Octave script \texttt{FE.m} requires values for $M$ and $\nu$ to be input. It first determines the coefficients $a_n$ in Eq.~\eqref{e-an}, and thus constructs initial data at $x=x_0$ for Eq.~\eqref{e-deV}; this process uses the scripts \texttt{find\_b.m, findJ\_0\_1\_2\_3.m, sumh.m, findBn.m}. Numerical integration then gives $J_{2T}(x)$ leading to $f_J(r),f_U(r),f_W(r)$, using \texttt{dJ2.m, int\_dx.m, FindRicci.m}. Then Eqs.~\eqref{e-C3050} are solved to find the coeffcients $C_{30},C_{50}$ so that $J(r),U(r),W_c(r)$ can be constructed. Finally, the script \texttt{find\_v1.m} evaluates $V_0^{[2,2]}(r), V_{ang}^{[2,2]}(r)$ and solves Eq.~\eqref{e-dV1} numerically to find $V_1^{[2,2]}(r)$; then \texttt{FindAve.m} evaluates $f_E(r,\nu,M)$ and plots $f_E(r,\nu,M)/f_E(r,\nu,0)$. When \texttt{FE.m} has completed, the user may also run \texttt{ChecksMink.m} (if $M=0$) to obtain the results for Table~\ref{t-Mink}, and/or \texttt{ChecksGen.m} to obtain the results in Table~\ref{t-Gen}. The Matlab/Octave script \texttt{Driver\_Fig1.m} runs \texttt{FE.m} multiple times with $M=1$ and various values of $\nu$ to produce Fig.~\ref{f-TI}. The script \texttt{diffsMinkSch.m} finds the change between using a Schwarzschild or Minkowski background in the values of $V^{[2,2]}_{1}$ and metric quantities reported in Sec.~\ref{s-conclusion}.
	
	\bibliographystyle{spphys}
	\bibliography{t_1,aeireferences}
\end{document}